# A Convolutional Neural Network-based Ensemble Post-processing with Data Augmentation for Tropical Cyclone Precipitation Forecasts

Sing-Wen Chen[1], Joyce Juang[2], Charlotte Wang[1,3], Hui-Ling Chang[2], Jing-Shan Hong[2], Chuhsing Kate Hsiao[1,3*]

[1]Institute of Health Data Analytics and Statistics, College of Public Health, National Taiwan University, Taiwan.
[2]Central Weather Administration, Taiwan.
[3]Master Program of Public Health, College of Public Health, National Taiwan University, Taiwan.

*Corresponding author, No. 17, Xu-Zhou Road, Taipei 10055, Taiwan. Email: ckhsiao@ntu.edu.tw

**Graphical abstract**

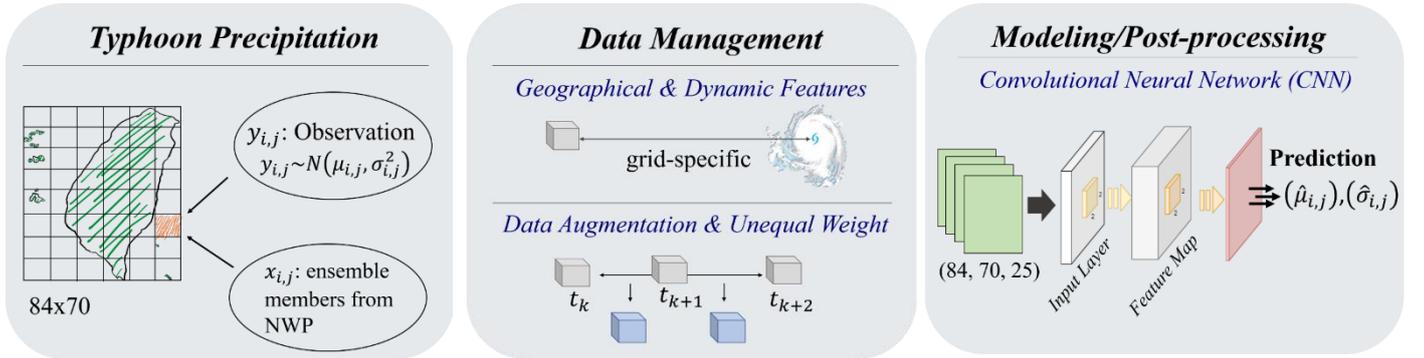

**Abstract**

    Heavy precipitation from tropical cyclones (TCs) may result in disasters, such as floods and landslides, leading to substantial economic damage and loss of life. Prediction of TC precipitation based on ensemble post-processing procedures using machine learning (ML) approaches has received considerable attention for its flexibility in modeling and its computational power in managing complex models. However, when applying ML techniques to TC precipitation for a specific area, the available observation data are typically insufficient for comprehensive training, validation, and testing of the ML model, primarily due to the rapid movement of TCs. We propose to use the convolutional neural network (CNN) as a deep ML model to leverage the spatial information of precipitation. The proposed model has three

distinct features that differentiate it from traditional CNNs applied in meteorology. First, it utilizes data augmentation to alleviate challenges posed by the small sample size. Second, it contains geographical and dynamic variables to account for area-specific features and the relative distance between the study area and the moving TC. Third, it applies unequal weights to accommodate the temporal structure in the training data when calculating the objective function. The proposed CNN-all model is then illustrated with the TC Soudelor's impact on Taiwan. Soudelor was the strongest TC of the 2015 Pacific typhoon season. The results show that the inclusion of augmented data and dynamic variables improves the prediction of heavy precipitation. The proposed CNN-all outperforms traditional CNN models, based on the continuous probability skill score (CRPSS), probability plots, and reliability diagram. The proposed model has the potential to be utilized in a wide range of meteorological studies.

1. **Introduction**

    Ensemble forecasts derived from Weather and Research Forecast (WRF) models are often generated based on different initial conditions, model parameters, and physical model systems. To quantify the inherent uncertainty in these forecasts, statistical post-processing procedures are utilized to improve forecast reliability and to provide probabilistic forecasts. Among these procedures, the Ensemble Model Output Statistics (EMOS) procedure and variants of Bayesian Model Averaging (BMA) have achieved successful applications in a wide range of weather variable predictions, including temperature (Gneiting et al., 2005), precipitation (Scheuerer, 2014; Huang et al., 2024), and wind speed (Edie et al., 2017; Pantillon et al., 2018), to name a few. These parametric models can provide probabilistic forecasts, but they are constrained by the requirement that the number of parameters must be smaller than the size of available data. Such difficulty may be addressed by artificial intelligence and machine learning (ML) approaches. Examples include random forests (Herman and Schumacher, 2018), neural networks (NNs), and deep learning (DL) neural networks; see McGovern et al. (2019) and Haupt et al. (2021) for more examples and reviews.

    Algorithms in the ML group have the advantage of integrating additional information, such as station-specific information and various meteorological variables, that cannot fit directly into the parametric distributional function. Specifically, the flexible setting of NNs can incorporate a large number of variables in

the input layer as predictors and accommodate nonlinear relationships between the input variables from different sources and the output variables, usually the parameters in the distributional function. Several studies have demonstrated the effectiveness of NNs and their beneficial contribution in applications. For instance, Rasp and Lerch (2018) illustrated a fully connected network (FCN) for ensemble post-processing to perform temperature forecasts in Germany. The output layer contained the predicted area-specific mean and standard deviation parameters of the normal distribution for temperature, and the input layer contained the mean and standard deviation of forecast members from numerical weather prediction (NWP) models as well as other auxiliary variables. Bremnes (2020) used a neural network to predict the Bernstein polynomial coefficients, adopted these output coefficients in a quantile function regression model, and showed its advantages in forecasting extreme wind speed in Norwegian synoptic stations. Later with the rise of deep learning algorithms, Grönquist et al. (2021) implemented a deep learning neural network, the convolutional neural network (CNN), with variables such as humidity and the U and V components of wind as input features to forecast the global temperature, and showed that their DL neural network could improve accuracy with less computational cost. In a Nature review article, Bauer et al. (2023) suggested that future weather and climate predictions can benefit from the use of artificial intelligence deep learning models.

The deep learning network CNN contains three types of layers: convolutional, pooling, and dense layers, in addition to the input and output layers. Various convolutional layers can filter out different features from the original input. Specifically, these convolutional layers extract spatial information from the input data, where the information can be linear or non-linear (Grönquist et al., 2021). This advantage often provides better performance in deterministic prediction or statistical post-processing, as compared with previous methods (Haupt et al., 2021). Several studies have adopted CNN in meteorology research. For instance, a study of atmospheric rivers (Chapman et al., 2019) used CNN to output a deterministic estimate of integrated vapor transport (IVT), Gagne et al. (2019) estimated directly the probability of severe hailstorms, Lagerquist et al. (2019) estimated the class probability of cold/warm/no front, and Shi et al. (2024) used CNN to predict extreme precipitation events. Based on a foundation of statistical post-processing with uncertainty quantification, two studies utilized CNN to estimate the parameter values: Grönquist et al. (2021) for global temperature and extreme events under the Gaussian distribution and Li et al. (2022) for summer precipitation under the Gamma distribution.

In contrast to the studies discussed above where it is straightforward to identify an extensive training dataset to train the CNN model, when the focus is on tropical cyclone (TC) precipitation in a specific area, the major challenge is the insufficient data for training the ML model due to the need for predictions within a relatively short timeframe. Tropical cyclones typically move forward at a speed faster than 63 kilometers per hour with heavy rain that can cause floods, landslides, and economic loss. The inference of heavy rain is critical for almost all TCs. The prediction of TC precipitation in any local region at time $T$ depends on the forecast-observation paired data prior to this time point. If the sample size is small, it will certainly compromise the performance of the deep learning-based algorithms. In recent medical image classification studies with CNN, data augmentation techniques have been considered to address the issue of insufficient data.

Various data augmentation tools have been used in medical image analysis to increase data sets, reduce class imbalance, expand data diversity, prevent overfitting, and enhance the generalizability of the model (Wang and Perez 2017; Mikołajczyk and Grochowski, 2018; Shorten and Khoshgoftaar, 2019; Chlap et al. 2021) in applications of CNN. These techniques may include geometric transformations, rotation, flipping, size rescaling, and mixing of images. While some of these are useful, some may not be suitable to train models for TC precipitation. For instance, the flips can produce additional data sets that are still useful and true in class labels for the whole image, but these flips may not represent reasonable or possible precipitation amounts in finer grids in the image. Therefore, only transformation and noise injection (Akbiyik, 2023) are considered in the current study as data augmentation tools in learning and training the CNN model for TC precipitation post-processing.

To account for the spatial structure of precipitation, this research adopts CNN as a DL model in the post-processing procedure. Specifically, the objective of this research is three-fold. First, we demonstrate the importance of data augmentation in the post-processing procedure for TC precipitation. The case study we considered here is the TC Soudelor that struck Taiwan in 2015. Taiwan island extends approximately 394 kilometers from north to south and about 144 kilometers from east to west. Often a given TC impacts Taiwan for less than 48 hours, while the ensemble forecast members produced by Taiwan Central Weather Administration at that time were recorded every six hours per day. Therefore, the size of the training and testing data sets can be as small as two in the deep learning model. This study will demonstrate the advantages of augmented data. Second, we incorporate dynamic features such as the distance between the study region and the center of the TC, which changes across time as the TC moves at a high speed. This distance is associated with the amount of precipitation and should be included in the input layers in the training model. Third, since the training data are the paired forecast-observation data collected longitudinally, they have temporal structures, some having been collected seconds before the current time point and others many hours before. Unequal weights in the loss function should therefore be assigned to the different components of the training data (Juang et al., 2022).

## 2. Materials and methods
### 2.1 Case study and data preparation

The study area is a rectangle encompassing Taiwan's islands (21.375°-25.525°N and 119.55°-123°E). This rectangle was gridded at the 0.05° ×0.05° resolution, resulting in a total of 5,880 (84×70) grids (Figure 1). Among them, 1,294 grids are on land (1,282 on Taiwan's main island and 12 on its outlying islands), while the rest are on the sea. The 1,294 land grids constitute the study region, including 623 plain grids and 671 mountain grids.

The time a tropical cyclone takes to pass through Taiwan is 1-2 days (Jian et al., 2022). The tropical cyclone considered here is Typhoon Soudelor, which travelled through Taiwan on August 7 and 8 in 2015. Its maximum fixed-time intensity reported by the Joint Typhoon Warning Center was 249 kph (155 mph) and it was classified as a super typhoon (Sopko & Falvey, 2015). Its maximum intensity near the typhoon's center was 48 meters per second when approaching Taiwan (CWA, 2015). The damage and loss of life it caused in the Northern Mariana Islands, Taiwan, and China were so catastrophic that the name was officially retired by the Economic and Social Commission for Asia and the Pacific and the World Meteorological Organization (ESCAP/WMO) Typhoon Committee in 2016 (ESCAP/WMO TC, 2016).

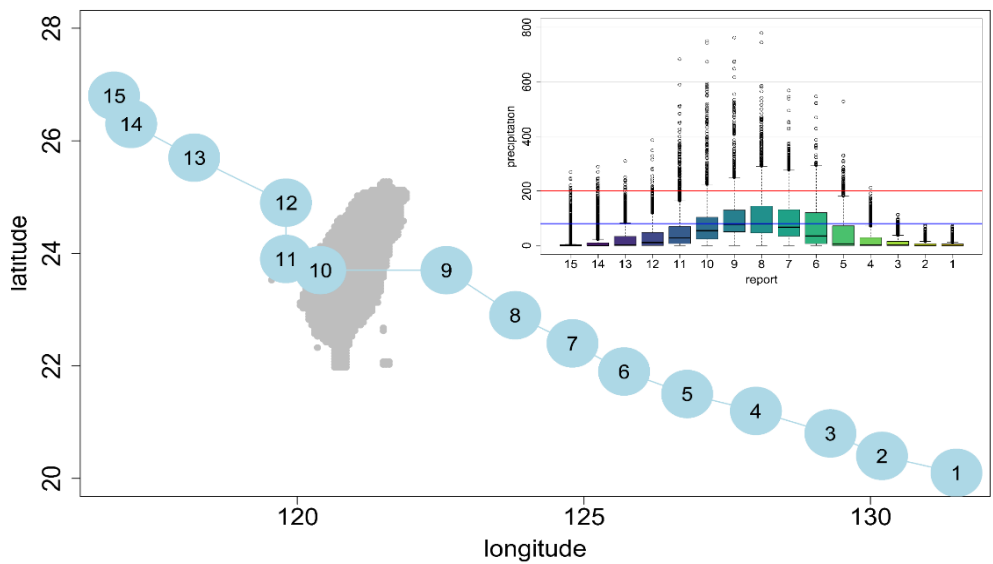

**Figure 1**. Study area, track of tropical cyclone Soudelor, and observed precipitation at different Reports.

The cumulative precipitation at each grid was recorded from 1800 UTC on August 5 to 0600 UTC on August 9, at a frequency of four times per day at 00, 06, 12, and 18 UTC. The track of this tropical cyclone is indicated with the curve in Figure 1 where numbers on the curve represent the longitudinal order of the precipitation record, referred to as the Report number, hereafter. A total of 15 Reports were collected.

The boxplots in Figure 1 show the distributions of the 24-hour accumulated precipitation, denoted as $y$, across the grids per Report. It is clear that heavy rain (if $y \in (80, 200]$mm) and beyond (if $y > 200$mm) were observed in more grids as the typhoon approached. These precipitation amounts, considered as the "true" observations, were derived from radar-estimated analysis using quantitative precipitation estimation (QPE) and Segregation Using Multiple Sensors (QPESUMS; Gourley et al., 2001). The horizontal resolution of the radar QPE from the QPESUMS was 1.25 km, but the grid unit in the current study was chosen to be 5 km to match the resolution of forecast member values. Every observed precipitation at each grid was associated with the twenty forecast member values that were generated from the WRF Ensemble Prediction System (WEPS) developed by the Central Weather Administration (CWA) in Taiwan (Li and Hong, 2011) with twenty sets of different model physics parameterizations, also recorded at the six-hour intervals.

**2.2 Geographical and dynamic features**

Grid-specific features were considered as input variables in the CNN model, including the longitude, latitude, and altitude of each grid, as terrain features often influence precipitation (Yeh and Elsberry, 1993; Su et al., 2012). Additionally, the track of the tropical cyclone and whether it has passed through the neighborhood can provide information on the amount of precipitation. For instance, the distance between the grid and the moving tropical cyclone is associated with precipitation. These features, in terms of longitude and latitude, are dynamic and should be included in the machine learning training model. Therefore, the input layers of the CNN model contain twenty forecast member values as well as the five geographical and dynamic features discussed above.

**2.3 Data augmentation**

Common data augmentation techniques such as flipping, rotation, and zooming are not appropriate in our case since the focus is on forecasting at each grid rather than a classification task of the whole image. Therefore, techniques that do not retain the grid-specific information are not considered here. In the current analysis, we adopt two types of augmented data. The first is the linear combination between two values from two consecutive reports. This was applied to each member forecast and the dynamic features. This procedure creates N-1 additional reports, in addition to the original N Reports. The second type of augmentation is noise injection to each of the above 2N-1 reports and leads to a total of 2×(2N-1) reports in the augmented report data. For instance, if the original reports are indexed with {1, 2, …, 15}, then the transformation creates reports indexed with {1.5, 2.5, …, 14.5} and the noise injection produces reports with indexes {1n, 1.5n, 2n, 2.5n, …, 14.5n, 15n}. Therefore, the final index set would be the union of {1, 1.5, 2, 2.5, …, 14.5, 15} and {1n, 1.5n, 2n, 2.5n, …, 14.5n, 15n}. At each original report k, i.e., k=1, 2, …, N, the data sets with index smaller than k are used to train the CNN post-processing model.

**2.4 Loss function and CNN model**

The CNN architecture for post-processing is shown in Figure 2. The input layers include the forecast members, geographical variables, and dynamic features, a total of 25 predictors over the 84×70 gridded area. The next convolutional procedure with 2×2 kernels across the input grids produces 32 feature maps. The final step produces two layers of output: one contains the estimated mean ($\hat{\mu}$) and the other the standard deviation ($\hat{\sigma}$) per grid. Both the design of output layer and the Gaussian distributional assumption follow the settings in Rasp and Lerch (2018).

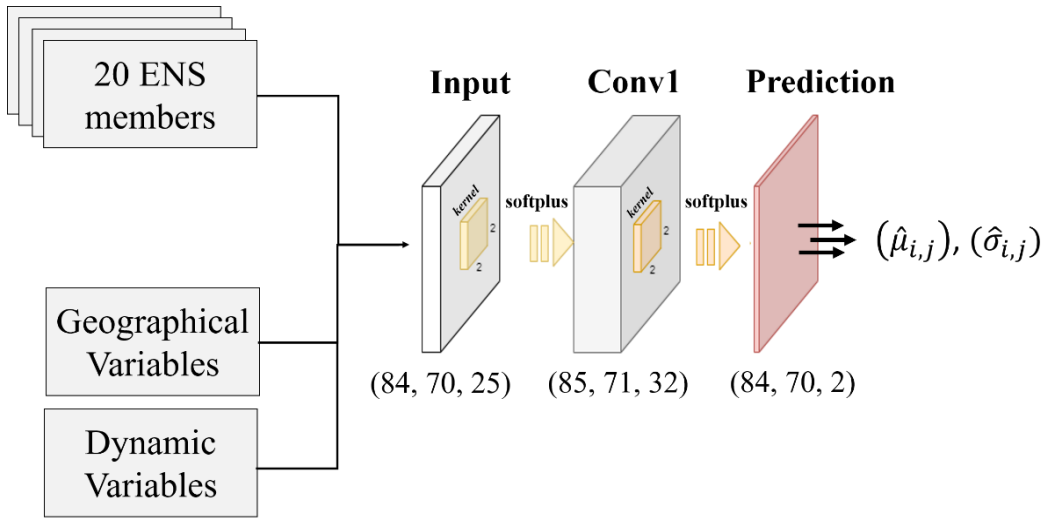

**Figure 2**. The proposed CNN models.

The Kaiming initialization was used to generate the initial weight in each layer and the softplus function was used as the activation function. The number of epochs was set at 100 and the learning rate at 0.001. The loss function was the continuous rank probability score (CRPS), defined as the distance between the predictive cumulative distribution function (CDF) and the observed precipitation,

$$\text{CRPS}(F, y) = \int_{-\infty}^{\infty} [F(q) - F_y(q)]^2 dq$$

Here $F(\cdot)$ is the grid-specific predictive CDF of the forecast distribution and $F_y$ is the CDF of the true observation $y$ (equivalent to an indicator function where $F_y(q) = 1$ if $q - y > 0$ and 0 otherwise). Based on the derivation in Gneiting et al. (2005), this CRPS function has a closed-form if $F$ corresponds to the CDF of $N(\mu, \sigma^2)$, and $\Phi$ and $\varphi$ are the CDF and the probability density function of a standard normal distribution, respectively.

$$\text{CRPS} = \sigma\left(\frac{y-\mu}{\sigma}\left(2\Phi\left(\frac{y-\mu}{\sigma}\right) - 1\right) + 2\varphi\left(\frac{y-\mu}{\sigma}\right) - \frac{1}{\sqrt{\pi}}\right)$$

This analytical form is utilized in the process of the CNN model optimization and the grid-specific mean $\mu$ and variance $\sigma^2$ are derived from the CNN output layer. All analyses were conducted in python v.3.9.12 and pytorch v.1.12.1.

The loss function CRPS is computed for each report in the training data. Since there are several reports in the training data, weights are assigned to different training reports to account for the temporal structure and relative contributions. The weights selected here follow the recommendation in Juang et al. (2022) where 1:2:4 is used if there are three training reports and 1:1:2:2:4:4 is used when augmented reports are included.

## 3  Results

A total of five models are demonstrated and compared, including the fully connected neural network (FCN) proposed by Rasp and Lerch (2018) with unequal weights as a non-deep learning model and four CNN models. The first CNN model, denoted as CNN in the following figures, considers only the twenty ensemble members in the traditional CNN. The second model adds in the five geographical and dynamic features, denoted as CNN-dyn. The third model contains the ensemble members and augmentation data, denoted as CNN-aug. The fourth model is the largest one, containing geographical and dynamic features, augmented data, and the ensemble forecasts, denoted as CNN-all. The differences are summarized in Table 1.

For TC Soudelor, there were fifteen original reports. The grid-specific observed precipitation amounts per report are classified into four categories in Table 2, very light rain if the cumulative amount $y \leq 10$mm, light rain if $y \in (10, 80]$mm, heavy rain if $y \in (80, 200]$mm, and beyond heavy rain if $y > 200$mm. The last group contains the categories of extremely heavy rain (if $y \in (200, 350]$mm) and torrential rain (if $y > 350$mm) since the number of grids with torrential rain is small. This classification follows the Hazard Condition defined by the Taiwan Central Weather Administration (https://www.cwa.gov.tw/V8/E/P/Warning/W26. html). Note that most areas, either plain or mountain, encountered heavy and extremely heavy rain in reports 7-10 (Table 2). We include two more reports to extend this set and demonstrate the results of the proposed post-processing for reports 6-11.

**Table 1.** Summary of the FCN model and the four CNN models.

| Models | Ensemble members | Geographical features | Dynamic features | Data augmentation |
|---|---|---|---|---|
| FCN | Y | – | – | – |
| CNN | Y | – | – | – |
| CNN-dyn | Y | Y | Y | – |
| CNN-aug | Y | – | – | Y |
| CNN-all | Y | Y | Y | Y |

**Table 2.** Values in parentheses are numbers of grids (left for plain, right for mountain) in each precipitation class (cumulative amount) per each forecast report.

| Index of Report | Very light rain, ≤10 mm | Light rain, (10, 80] mm | Heavy rain, (80, 200] mm | Beyond heavy rain, >200 mm |
|---|---|---|---|---|
| 1 | (543, 566) | (80, 105) | (0, 0) | (0, 0) |
| 2 | (546, 563) | (77, 108) | (0, 0) | (0, 0) |
| 3 | (511, 488) | (107, 168) | (5, 15) | (0, 0) |
| 4 | (487, 435) | (117, 178) | (17, 53) | (2, 5) |
| 5 | (384, 240) | (177, 255) | (49, 109) | (13, 67) |
| 6 | (84, 45) | (355, 307) | (130, 156) | (54, 163) |
| 7 | (14, 2) | (299, 166) | (238, 250) | (72, 253) |
| 8 | (1, 0) | (186, 117) | (313, 220) | (123, 334) |
| 9 | (1, 1) | (212, 167) | (243, 195) | (167, 308) |
| 10 | (35, 14) | (266, 304) | (176, 181) | (146, 172) |
| 11 | (108, 99) | (217, 348) | (220, 94) | (78, 130) |
| 12 | (144, 148) | (295, 360) | (178, 134) | (6, 29) |
| 13 | (216, 258) | (331, 324) | (75, 84) | (1, 5) |
| 14 | (346, 522) | (262, 136) | (14, 10) | (1, 3) |
| 15 | (489, 593) | (133, 67) | (0, 8) | (1, 3) |

**3.1 Evaluation and comparison with CRPSS**

With the focus on heavy rain and beyond, we demonstrate the distributions of the continuous probability skill score (CRPSS) for Reports 6-11 in Figure 3 using the reference of a Gaussian distribution with the ensemble mean and variance as the parameter values. Figure 3 contains the skill scores for plain grids if the altitude is below 500 meters, while those for mountain grids are shown in Supplementary Figure S1.

Note that the CRPSS under the four CNN models is mostly positive, indicating a better fit than the ensemble model. Among the five models (one FCN and four CNNs), CNN-all performs the best, as indicated by its relatively larger positive CRPSS values with a smaller variation. Incorporating the augmented data and variables of geographical and dynamic features improves the post-processing procedure with the ensemble members. The four CNN models perform better at the plain grids but not as good as the FCN model at the mountain grids in some Reports. This is because the FCN model tends to provide a larger prediction probability of a certain amount of precipitation compared to the other models, regardless of the amount. This will become clear in the next section when examining the probability plot. For light rain and very light rain, the Gaussian distribution based on the ensemble mean and variance provides a better post-processing procedure, as indicated in Supplementary Figure S2.

**3.2 Evaluation and comparison with probability plot and reliability diagram**

The chance of disaster increases when the precipitation amount is large. Here we utilize the advantage of the statistical post-processing procedure to examine the probabilistic prediction. In Figures 4 and 5, the original precipitation amounts, classified by three intervals ($\geq 80$, between 200 and 350, and $\geq 350$}, were compared against the prediction probability of precipitation larger than 200mm. Only the grids with probabilities larger than 0.5 are indicated. Depending on the magnitude of the probability, the areas are colored with light to dark colors. These probabilities were calculated under the FCN and CNN-all models in previous figures and the traditional ensemble member forecast model, denoted as Members.

For grids with a true precipitation amount reaching 200mm (indicated by the two darker shades of blue dots in the top panels in Figures 4 and 5), all three models provide large prediction probabilities. However, for grids with precipitation less than 200mm (light blue dots in the top panels in Figures 4 and 5), both the Members model and the FCN model are more conservative by delivering probability values larger than 0.9 at these grids. In general, the spatial pattern in the observed precipitation amounts is better captured by the CNN-all model.

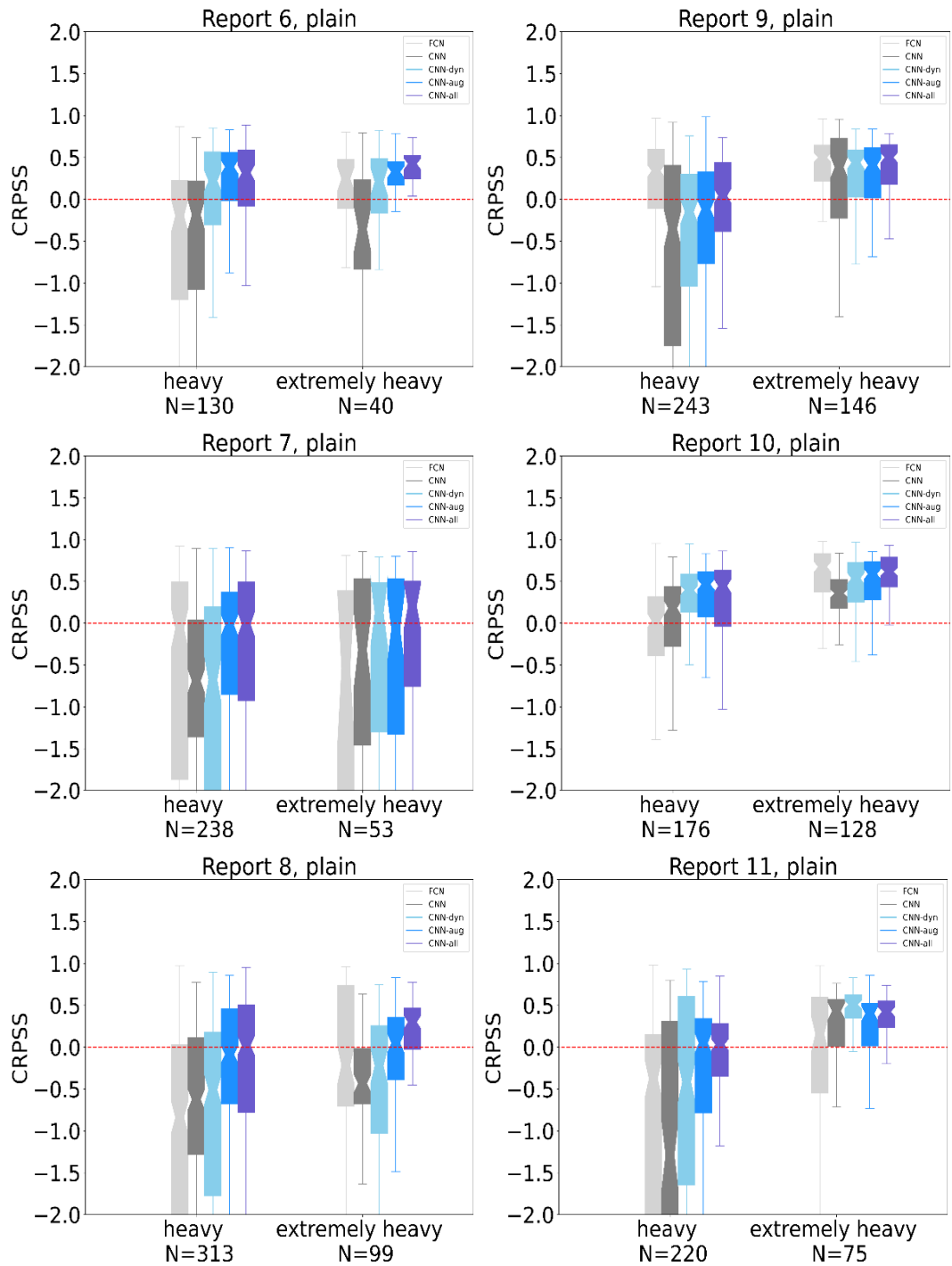

**Figure 3**. Boxplots of CRPSS under categories of heavy rain and beyond in plain grids for Reports 6-11.

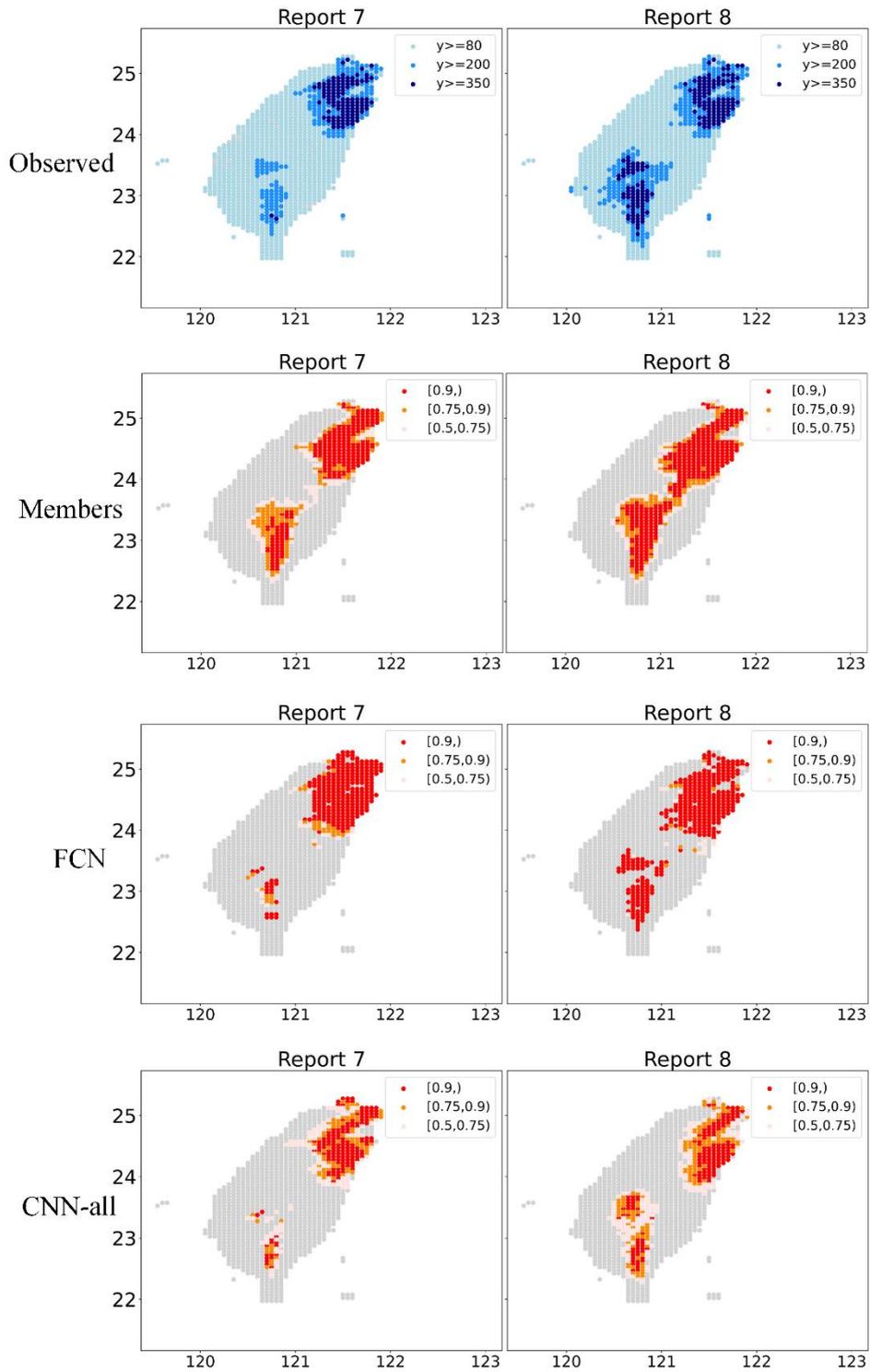

**Figure 4**. Probability of precipitation larger than 200mm per grid for Reports 7-8.

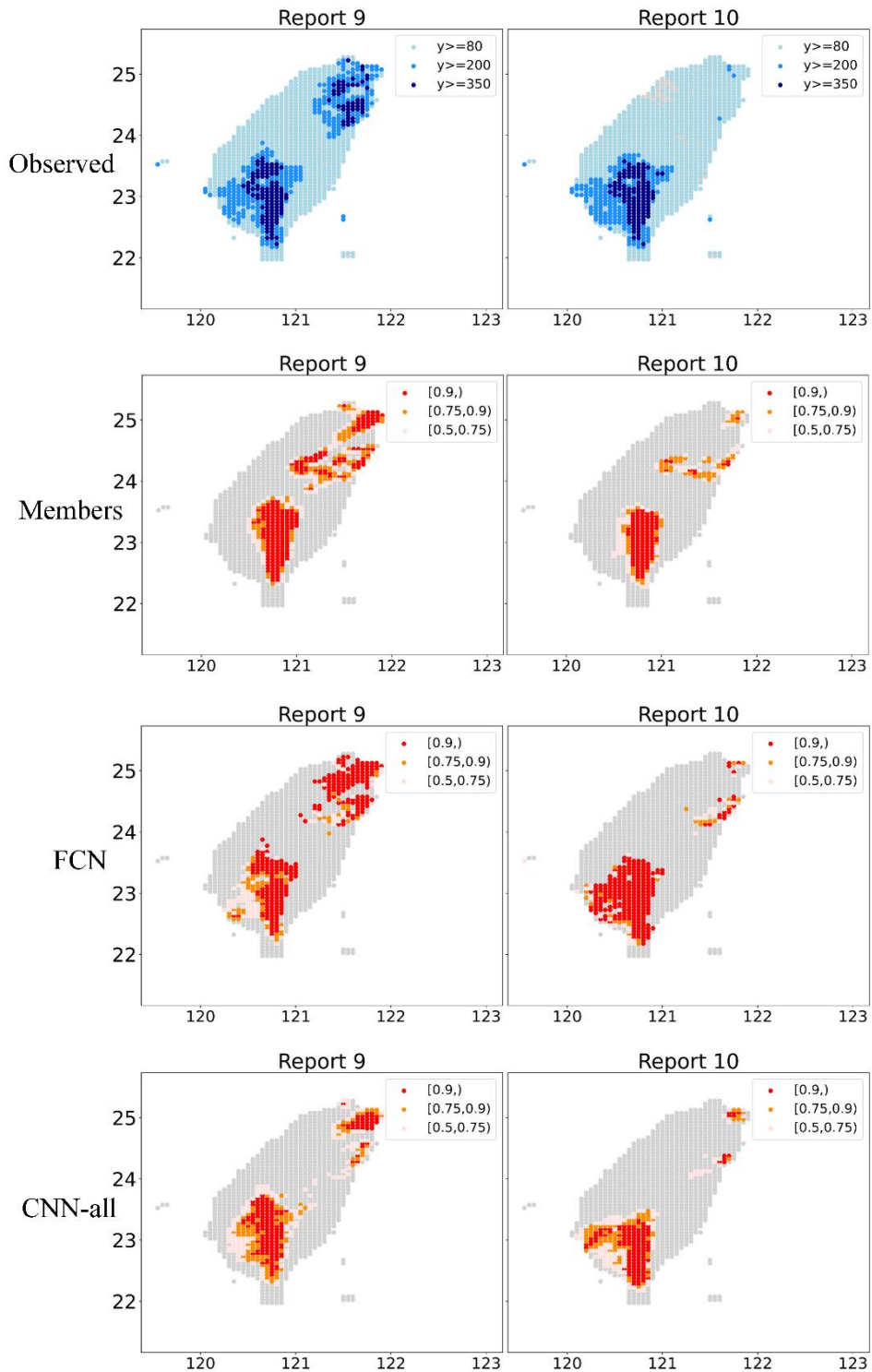

**Figure 5**. Probability of precipitation larger than 200mm per grid for Reports 9-10.--

When comparing the probability prediction with true observed precipitations exceeding 200mm, CNN-all provides better performance than the ensemble members and FCN, as demonstrated in the reliability diagram in Figure 6. In Reports 7-10, a total of 325, 457, 475, and 318 plain and mountain grids exhibited cumulative precipitation over 200mm. In each panel in Figure 6, the CNN-all post-processing model exhibits the closest curve to the observed frequencies. In contrast, both the ensemble members and FCN tend to predict with a probability larger than the observed frequency.

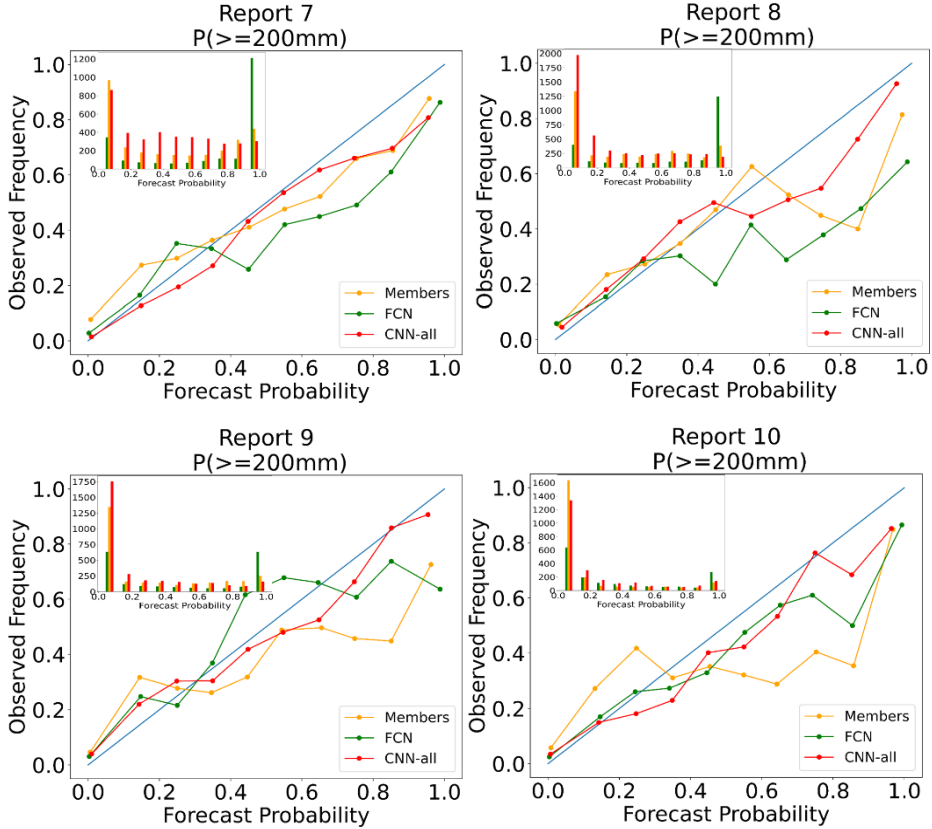

**Figure 6**. Reliability diagram for precipitation beyond 200mm.

## 4 Discussion

This study aims to build a CNN-based post-processing ensemble forecast model for tropical cyclone precipitation. Specifically, we applied the data augmentation techniques to enhance the CNN learning, as TCs typically move at a fast speed and provide limited data for training the DL model. We also incorporated the geographical and dynamic variables to increase the performance of the post-processing procedure. The case study of Typhoon Soudelor's impact on Taiwan demonstrates that the proposed CNN-all model performs better compared to other DL algorithms.

The proposed CNN with data augmentation algorithm for post-processing can be easily generalized to incorporate other factors impacting the precipitation as features, including temperature, pressure, wind speed, sea level pressures (Shi et al, 2024), and spatial structure of TCs (Qin et al., 2024). The incorporation of these features in the input layer of the DL model is straightforward and may boost the model performance.

This flexibility is an advantage that traditional statistical parametric models do not share.

Alternative data augmentation tools are available, which can be tailored to specific domains. For instance, if the whole study region is to be labeled for precipitation exceeding a pre-determined value, then each map of the member-specific forecast can be considered as an image and be applied with common augmentation tools for image analysis, including transformations, rotation, and flipping. Such augmentation may change the grid content but will not distort the original whole image to something completely different. It is noted that these methods are not proper for the current study because our study interest does not lie in the prediction of summary information for the whole map, but in forecasting the grid-specific parameters. Therefore, it is recommended to retain the spatial relationship of the grids and spatial structure of TCs when designing the augmentation tools, as Maxwell et al. (2021) and Qin et al. (2024) suggested in their investigations of the impact of recent climate change on TC precipitation extremes.

Several issues in the case study merit attention. First, CNN-all was not the best model across Reports in the post-processing forecasts in mountain areas, especially for Report 9 (Supplementary Figure S1), whose corresponding forecast was for the time interval in which Soudelor made landfall over Hualian at 20:40 UTC and was weakened when it encountered the Central Mountains. In other words, the significantly varying altitudes of Taiwan's mountains and the terrain-locking effect (Chang et al., 1993; Su et al., 2012) considerably affected the structure of Soudelor and hence influenced its precipitation. Additionally, the smooth kernels in CNN may improperly level out the spatial effect and overlook the influence of the terrain-locking effect when the spatial structure changes dramatically in neighboring grids. One solution to this issue would be to design a grid-specific post-processing model. Future work is warranted to incorporate this decision into the proposed CNN. Second, variables in addition to the geographical and dynamic variables considered in the current study may improve the model performance. For instance, Qin et al. (2024) proposed a metric, DIST30, the mean radial distance from centers of clustered heavy rainfall cells to the TC center, as a variable to model TC precipitation. Information of historical TCs traveling in similar paths may provide insight as well (Chang et al. 1993). Third, the case study aimed to forecast cumulative precipitation in the coming twenty-four hours. However, in practice, the true observations would not have been available until one day later. A better demonstration of the proposed model would be cases where the accumulation period is shorter so that when training the model in practice, both the input and outcome data are ready and we are still in time to make real-time forecasts. Finally, the case study considered here adopted 6-hourly ensemble forecasts. If these forecasts can be made over a shorter interval, say per 3 hours, then the number of training data sets can be doubled and perhaps better meet the large data requirement for DL models. A generative AI *CorrDiff* was recently proposed to provide a global-to-km scale downscaling model (Mardani et al., 2023). It would be interesting to investigate if this data-driven model can be applied to obtain finer scales temporally. However, before this can be achieved and the cost for computation can be further lowered, CNN with data augmentation remains a competitive choice for post-processing ensemble forecasts.

## 5  Conclusions

This study adopted CNN to consider the spatial structure of tropical cyclone precipitation and used unequal weights in training data to incorporate temporal information. We further included augmented data to ease the tension of insufficient training data as well as geographical and dynamic variables to expand the flexibility of the model. The proposed CNN-all model performed better in predicting heavy rain and beyond. The application of data augmentation techniques warrants attention in future studies.

**Declaration of competing interest**

The authors declare that they have no known competing financial interests or personal relationships that could have appeared to influence the work reported in this paper.

**Data availability**

The data can be applied via the Taiwan Central Weather Administration, Typhoon Database (https://opendata.cwa.gov.tw/index or https://rdc28.cwa.gov.tw/TDB/). The authors do not have permission to share the data.

**Appendix A. Supplementary data**

Supplementary data to this article can be found online. Figures S1 and S2.

**Supplementary Figure S1.** Boxplots of CRPSS under the FCN and four CNN models.

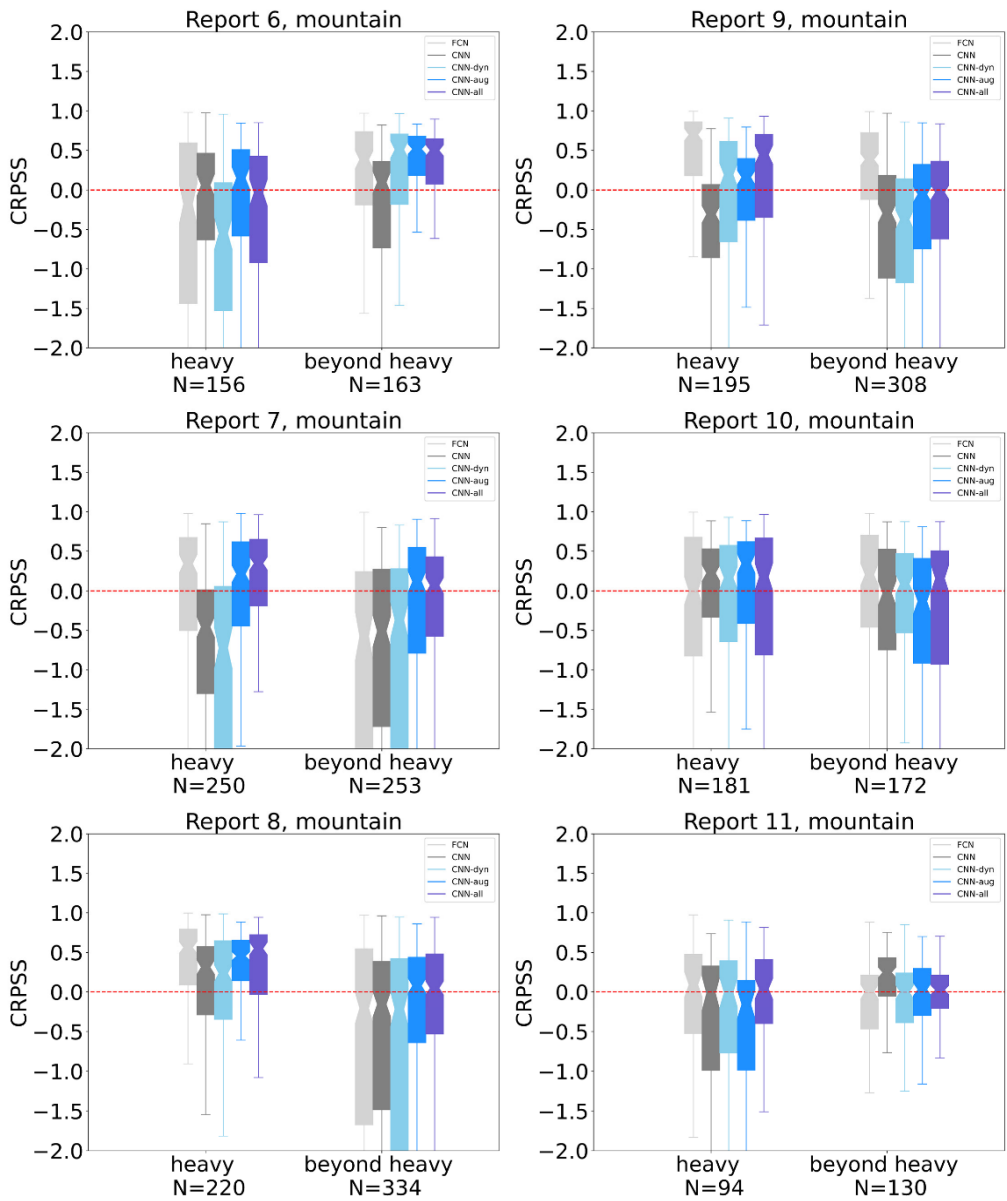

**Supplementary Figure S2.** Boxplots of CRPSS under the FCN and four CNN models.

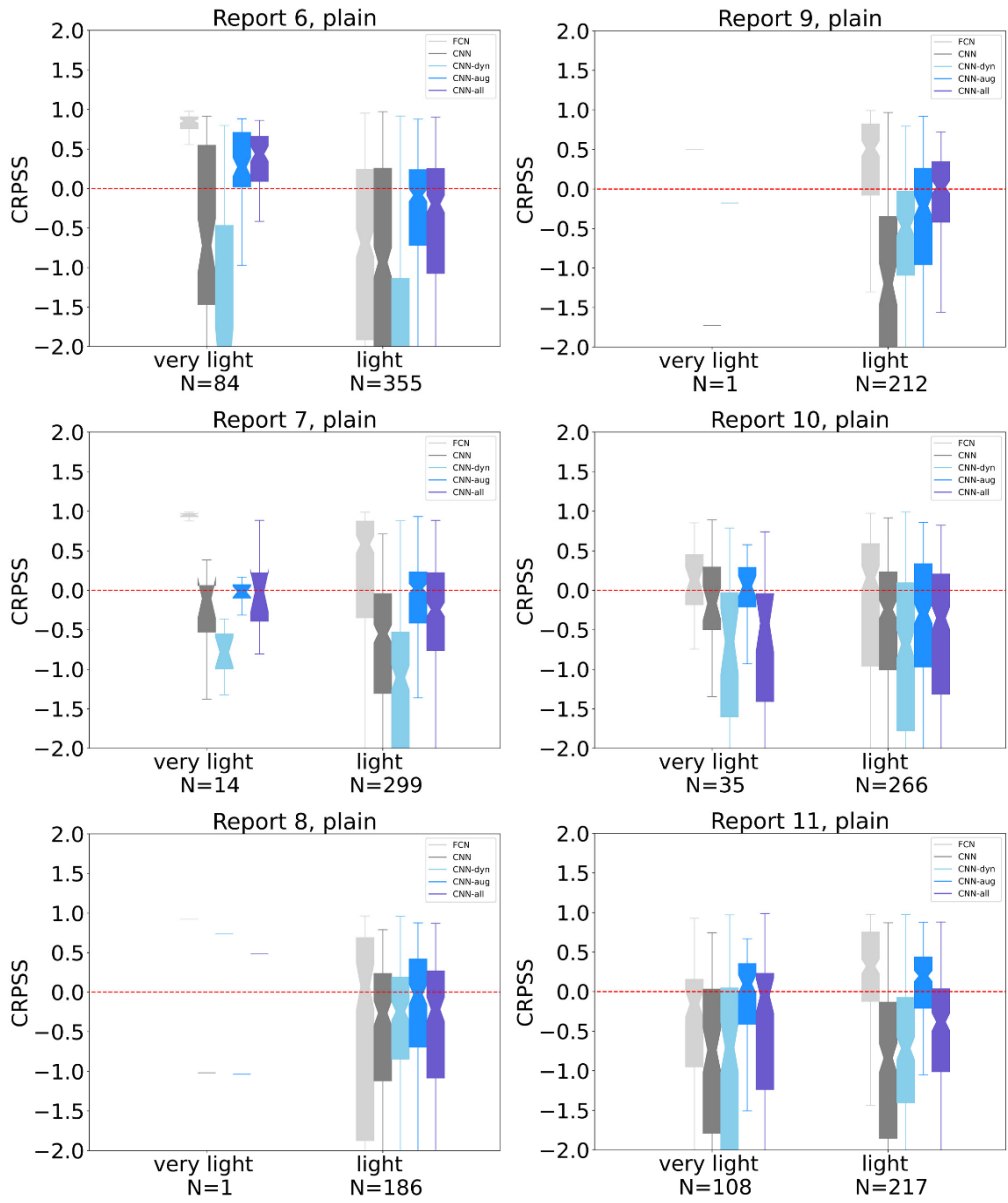